\documentclass[10pt]{article}
\usepackage{epsf} 
\usepackage{amsmath}
\usepackage{amssymb}
\usepackage{epsfig}
\usepackage{latexsym}
\usepackage{amsfonts}
\usepackage{graphicx}%
\usepackage{varioref}
\usepackage{ifthen}
\begin{document}
\parindent 0mm 
\setlength{\parskip}{\baselineskip} 
\pagenumbering{arabic} 
\setcounter{page}{1}
\mbox{ }
\rightline{UCT-TP-301/2014}
\newline
\newline
\rightline{August 2014}
\newline

\begin{center}
\Large \textbf{Analytical determination of the QCD quark masses}
{\footnote{{\LARGE {\footnotesize To appear as a chapter in the book {\it{Fifty Years of Quarks}}, H. Fritzsch and M. Gell-Mann, editors (World Scientific Publishing Company, Singapore). }}}}
\end{center}
\begin{center}
C. A. Dominguez
\end{center}

\begin{center}
Centre for Theoretical and Mathematical Physics  and Department of Physics, University of Cape Town, Rondebosch 7700, South Africa
\end{center}

\begin{center}
\textbf{Abstract}
\end{center}

\noindent
The current status of determinations of the QCD running quark masses is reviewed. Emphasis is on recent progress on analytical precision determinations based on finite energy QCD sum rules. A critical discussion of the merits of this approach over other alternative QCD sum rules is provided.  
Systematic uncertainties from both the hadronic and the QCD sector have been recently identified and dealt with successfully, thus leading to values of the quark masses with unprecedented accuracy.  Results currently rival in precision with lattice QCD determinations. 

\section{Introduction}
Quark and gluon confinement in Quantum Chromodynamics (QCD) precludes direct experimental measurements of the fundamental QCD parameters, i.e. the strong interaction coupling and the quark masses. Hence, in order to determine these parameters analytically one needs to relate them to experimentally measurable quantities. Alternatively,  simulations of QCD on a lattice (LQCD) provide increasingly accurate numerical values for these parameters, but little if any insight into their origin. The first approach relies on the intimate relation between QCD Green functions, in particular their Operator Product Expansion (OPE) beyond perturbation theory, and their hadronic counterparts. This relation follows from Cauchy's theorem in the complex energy plane, and is  known as the finite energy QCD sum rule (FESR) technique \cite{QCDSR}. In addition to producing numerical values for the QCD parameters, this method provides a detailed breakdown of the relative impact of the various dynamical contributions. For instance, the strong coupling at the scale of the $\tau$-lepton mass  essentially follows from the relation between the experimentally measured $\tau$ ratio, $R_\tau$, and a contour integral involving the perturbative QCD (PQCD) expression of the $V+A$ correlator, a classic example of a FESR. This is the cleanest, most transparent, and model independent determination of the strong coupling \cite{PICH1}-\cite{PICH2}. It also allows to gauge the impact of each individual term in PQCD, up to the currently known five-loop order.
Similarly, in the case of the quark masses one considers a QCD correlation function which on the one hand involves the quark masses and other QCD parameters, and on the other hand it involves a measurable (hadronic) spectral function. Using Cauchy's theorem to relate both representations, the quark masses  become a function of QCD parameters, e.g. the strong coupling, some vacuum condensates reflecting confinement, etc., and measurable hadronic parameters. The virtue of this approach is that it provides a breakdown of each contribution to the final value of the quark masses. More importantly, it allows to tune the relative weight of each of these contributions by introducing suitable integration kernels. This last feature has been used recently in the case of the charm- and bottom-quark masses leading to very accurate values. It has also been employed to unveil the hadronic systematic uncertainties affecting light quark mass determinations, to wit. In this case the ideal Green function is the light-quark pseudoscalar current correlator. This contains the square of the quark masses as an overall factor multiplying the PQCD expansion, and the leading power corrections in the OPE. Unfortunately, this correlator is not realistically accessible experimentally beyond the pseudoscalar meson pole. While the existence of at least two radial excitations
of the pion and the kaon are known from hadronic interaction data, this information is hardly enough to reconstruct the full  spectral functions. In spite of many attempts over the years to model them, there remained an  unknown systematic uncertainty that has plagued light quark mass determinations from QCD sum rules (QCDSR). The use of the vector current correlator, for which there is plenty of experimental data from $\tau$ decays and $e^+ e^-$ annihilation, is not a realistic option for the light quarks as their masses enter as sub-leading terms in the OPE. The scalar correlator, involving the square of quark mass differences, at some stage offered some promise  for determining the strange quark mass with reduced systematic uncertainties. This was due to the availability of data on $K-\pi$ phase shifts. Unfortunately, these data do not fully determine the hadronic spectral function. The latter can be reconstructed from phase shift data only after substantial theoretical manipulations, implying a large unknown systematic uncertainty. A breakthrough has been made recently  by introducing an integration kernel in the contour integral in the complex energy plane. This allows to suppress substantially the unknown hadronic resonance contribution to the pseudoscalar current correlator. As it follows from Cauchy's theorem, this suppression implies that the quark masses are determined essentially from the well known pseudoscalar meson pole and PQCD (well known up to five-loop level). In this way it has  been possible to reduce the hadronic resonance contribution to the 1\% level, allowing for an unprecedented accuracy of some $8 - 10\, \%$ in the values of the up-, down-, and strange-quark masses. Nevertheless, there still remained a well known shortcoming in the PQCD sector due to the poor convergence of the pseudoscalar correlator. In fact, the contribution to the quark mass from each perturbative term, up to five-loop level, is essentially identical. While this problem was well known, it remained unresolved for decades. A breakthrough has finally been made recently by using Pad\`{e} approximants to accelerate efficiently the perturbative convergence. When used for the strange-quark mass this procedure unveils a  systematic uncertainty of some 30\% in all previous determinations based on the original perturbative expansion. The strange-quark mass is now known with a 10\% error, but essentially free from systematics from the hadronic and the QCD sector. Further improvement on this accuracy will be possible with further reduction of the uncertainty in the strong coupling, now the main source of error.\\
The determination of the charm- and bottom-quark masses has been free of  systematic uncertainties due to the hadronic resonance sector, as there is plenty of experimental information in the vector channel from $e^+ e^-$ annihilation into hadrons. One problem, though, is that the massive vector current correlator is not known in PQCD to the same level as the light pseudoscalar correlation function. Nevertheless, substantial theoretical progress has been made over the years leading to extremely accurate charm- and bottom-quark masses. The novel idea of introducing suitable integration kernels in Cauchy's contour integrals, as described above, has also been used recently as a way of improving accuracy in the heavy-quark sector. For instance, kernels can be used to suppress regions where the data is either not as accurate, or simply unavailable. This will also be reported here.\\

The paper is organized as follows. First, determinations of quark-mass ratios from various hadronic data, as well as from chiral perturbation theory (CHPT), will be reviewed in Section 2. These ratios are quite useful as consistency checks for results from QCDSR. Section 3 describes the OPE beyond perturbation theory, one of the two pillars of QCDSR. Section 4 discusses quark-hadron duality and FESR, while Section 5 provides a critical discussion of Laplace sum rules, as originally proposed and applied to a large number of issues in hadronic physics. With precision determinations being the name of the game at present, these Laplace sum rules cannot compete in accuracy with e.g. FESR, thus falling out of favour. Several conceptual flaws affecting these Laplace sum rules are pointed out and discussed, paving the way for FESR as the preferred method to determine QCD as well as hadronic parameters. 
FESR weighted by suitable integration kernels will be analysed in the light quark sector in Section 6. In particular it will be shown how this technique unveils the underlying hadronic systematic uncertainty plaguing  light quark mass determinations for the past thirty years. 
In Section 7 recent progress on charm- and bottom-quark mass determinations will be reported. Comparison with LQCD results for all quark masses will also be made. Finally, Section 8  provides  a very brief summary of this report.\\

As an important disclaimer, this paper is not a comprehensive review of all past quark mass determinations from QCDSR. It is, rather, a report on very recent progress on the subject. Given that past determinations of light quark masses were affected by unknown systematic uncertainties, both from the hadronic resonance sector as well as the QCD sector, it makes little sense to review them. Any agreement between values affected by these  uncertainties and current results, free of them, would only be fortuitous. For instance, once the hadronic resonance uncertainty is removed, the values of all three light quark masses get reduced by some 15 - 20 \%, with a similar situation in the QCD sector (due to the solution of the well known problem with the poor convergence of the pseudoscalar correlator). This is a clear sign of a systematic uncertainty acting in only one direction.  In addition, light quark masses from QCDSR before 2006 employed correlators  up to at most four-loop level in PQCD, together with superseded values of the strong coupling. Last but not least, quark masses determined from Laplace QCDSR are affected by very large, mostly unacknowledged, systematic uncertainties from the hadronic as well as the QCD sector, as discussed in Section 5.
\section{Quark mass ratios}
Quark masses actually precede QCD by a number of years, albeit under the guise of {\it current algebra quark masses}, which clearly lacked today's detailed understanding of quark-mass renormalization. In fact, the study of global ${\mbox{SU}}(3) \times {\mbox{SU}}(3)$ chiral symmetry realized \'{a} la Nambu-Goldstone, and its breaking down to ${\mbox{SU}}(2) \times {\mbox{SU}}(2)$, followed by a breaking down to ${\mbox{SU}}(2)$, and finally to ${\mbox{U}}(1)$  was first done using the strong interaction Hamiltonian \cite{CA1}-\cite{HLR}
\begin{equation}
H(x)\, =\, H_0(x)\, + \,\epsilon_0\; u_0(x)\, + \,\epsilon_3\; u_3(x)\, + \,\epsilon_8\; u_8(x) \;.
\end{equation}
The term $H_0(x)$ above is $\mbox{SU}(3) \times \mbox{SU}(3)$ invariant, the $\epsilon_{0,3,8}$  are symmetry breaking parameters, and the scalar densities $u_{0,3,8}(x)$  transform according to the $3 \,\overline{3} \;+\;\overline{3} \,3$ representation of $\mbox{SU}(3) \times \mbox{SU}(3)$. In modern language, $\epsilon_8$ is related to the strange quark mass $m_s$, and $\epsilon_3$ to the difference between the down- and the up-quark masses $m_d-m_u$, while the scalar densities are related to products of quark-anti-quark field operators.  For instance, the ratio of $\mbox{SU}(3)$ breaking to $\mbox{SU}(2)$ breaking is given by
\begin{equation}
R \equiv \frac{m_s - m_{ud}}{m_d - m_u} = \frac{\sqrt{3}}{2} \; \frac{\epsilon_8}{\epsilon_3} \;,
\end{equation}
where $m_{ud} \equiv (m_u + m_d)/2$. In the pre-QCD era many relations for quark-mass ratios were obtained from hadron mass ratios, as well as from other hadronic information, e.g. $\eta \rightarrow 3 \pi$, $K_{l 3}$ decay, etc. \cite{HLR}.  To mention a pioneering  determination of the ratio $R$ above, from a solution to the $\eta \rightarrow 3 \pi$ puzzle proposed in  \cite{CAD1}  it followed \cite{CAD2}  $R^{-1} = 0.020 \pm 0.002$, in remarkable agreement with a later determination based on baryon mass splitting \cite{MZ} $R^{-1} = 0.021 \pm 0.003$, and with  the most recent value \cite{FLAGR} $R^{-1} = 0.025 \pm 0.003$ . With the advent of CHPT \cite{HP}-\cite{HLR}, \cite{FLAGR}-\cite{GC}, certain quark mass ratios turned out to be renormalization scale independent to leading order, and could be expressed in terms of pseudoscalar meson mass ratios \cite{HLR},\cite{CHPT1}-\cite{SW}, e.g.
\begin{equation}
\frac{m_u}{m_d} = \frac{M_{K^+}^2 - M_{K^0}^2 + 2 M_{\pi^0}^2 - M_{\pi^+}^2}{M_{K^0}^2 - M_{K^+}^2 + M_{\pi^+}^2} = 0.56 \;,
\end{equation}
\begin{equation}
\frac{m_s}{m_d} = \frac{M_{K^+}^2 + M_{K^0}^2 - M_{\pi^+}^2}{M_{K^0}^2 - M_{K^+}^2 + M_{\pi^+}^2} = 20.2 \;,
\end{equation}
where the numerical results follow after some subtle corrections due to electromagnetic self energies \cite{FLAGR}.
Beyond leading order in CHPT things become complicated. At next to leading order (NLO) the only  parameter free relation is
\begin{equation}
Q^2 \equiv \frac{m_s^2 - m_{ud}^2}{m_d^2 - m_u^2} =
\frac{M_{K}^2 - M_{\pi}^2}{M_{K^0}^2 - M_{K^+}^2}\; \frac{M_K^2}{M_{\pi}^2} \;. 
\end{equation}
Other quark mass ratios  at NLO and beyond depend on the renormalization scale, as well as on some CHPT low energy constants which need to be determined independently \cite{FLAGR}-\cite{CHPT1}. After taking into account electromagnetic self energies, Eq.(5) gives \cite{CHPT1}  $Q = 24.3$, a recent analysis of $\eta \rightarrow 3 \pi$ \cite{CHPT1}, \cite{GC} gives $Q = 22.3 \pm 0.8$, and the most recent value from the FLAG Collaboration \cite{FLAGR} is
\begin{equation}
Q = 22.6 \pm 0.7 \pm 0.6 \;.
\end{equation}
The ratios $R$, Eq.(2), and $Q$, Eq.(5), together with the leading order ratios Eqs.(3)-(4), will prove useful for comparisons with QCD sum rule results. An additional useful quark mass ratio involving the
ratios Eqs.(3)-(4) is
\begin{equation}
r_s \equiv \frac{m_s}{m_{ud}} = \frac{2\; m_s/m_d}{1 + m_u/m_d} = 28.1 \pm 1.3 \;,
\end{equation}
where the numerical value follows from the NLO CHPT relation \cite{CHPT1}, to be compared with the LO result from Eqs.(3)-(4), $r_s = 25.9$, and a large $N_c$ estimate \cite{eta} $r_s = 26.6 \pm 1.6$. The most recent FLAG Collaboration result is \cite{FLAGR}
\begin{equation}
r_s = 27.46 \pm 0.15 \pm 0.41 \;.
\end{equation}
\section{Operator product expansion beyond perturbation theory}
The OPE beyond perturbation theory in QCD, one of the two pillars of the sum rule technique, is an effective
tool to introduce quark-gluon confinement dynamics. It is not a model, but rather a parametrization of quark and gluon propagator corrections due to confinement, done in a rigorous renormalizable quantum field theory framework. Let us consider a typical object in QCD in the form of the two-point function, or current correlator
\begin{equation}
\Pi(q^2)\,=\,i\; \int \,d^4 x \; e^{iqx} \; <0|\,T(J(x)\,J(0))\,|0 >,
\end{equation}
where the local current $J(x)$ is built from the quark and gluon fields entering the QCD Lagrangian. Equivalently, this current can also be written in terms of hadronic fields with the same quantum numbers. A relation between the two representations follows from Cauchy's theorem in the complex energy (squared) plane. This is often referred to as quark-hadron duality, the second pillar of the QCDSR method to be discussed in the next section.
The QCD correlator, Eq.(9),  contains a perturbative piece (PQCD), and a non perturbative one mostly reflecting quark-gluon confinement. The leading order in PQCD is shown in Fig.1.  Since confinement has not been proven analytically in QCD, its effects  can only be introduced effectively, e.g. by parametrizing quark and gluon propagator corrections in terms of vacuum condensates. This is done as follows. In the case of the quark propagator
\begin{equation}
S_F (p) = \frac{i}{\not{p} - m}\;\;\Longrightarrow \;\;\frac{i}{\not{p} - m + \Sigma(p^2)} \;, 
\end{equation}
the  propagator correction $\Sigma(p^2)$  contains the information on confinement, a purely non-perturbative effect. One expects this correction to peak at and near the quark mass-shell, e.g. for $p \simeq 0$ in the case of light quarks. Effectively, this can be viewed as in Fig. 2, where the (infrared) quarks in the loop have zero momentum and interact strongly with the physical QCD vacuum. This effect is then parametrized in terms of the quark condensate $\langle 0| \bar{q}(0) q(0) | 0 \rangle$.
\begin{figure}[ht]
\begin{center}
  \includegraphics[height=.1\textheight]{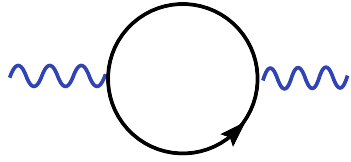}
  \caption{\footnotesize{
  Leading order PQCD correlator. All values of the four-momentum of the quarks in the loop are allowed. The  wiggly line represents the current of momentum $q$ ($-q^2 >> 0$).}}
  \label{fig:figure1}
\end{center}
\end{figure}
Similarly, in the case of the gluon propagator 
\begin{equation}
D_F (k) = \frac{i}{k^2}\;\;\Longrightarrow \;\;\frac{i}{k^2 + \Lambda(k^2)} \;,
\end{equation}
the propagator correction $\Lambda(k^2)$ will peak at $k\simeq 0$, and the effect of confinement in this case can be parametrized by the gluon condensate $\langle 0| \alpha_s\; \vec{G}^{\mu\nu} \,\cdot\, \vec{G}_{\mu\nu}|0\rangle$ (see Fig.3).
In addition to the quark and the gluon condensate there is a plethora of higher order condensates entering the OPE of the current correlator at short distances, i.e.
\begin{equation}
\Pi(q^2)|_{QCD}\,=\, C_0\,\hat{I} \,+\,\sum_{N=0}\;C_{2N+2}(q^2,\mu^2)\;\langle0|\hat{O}_{2N+2}(\mu^2)|0\rangle \;,
\end{equation}
where $\mu^2$ is the renormalization scale, and where the Wilson coefficients in this expansion, 
$ C_{2N+2}(q^2,\mu^2)$,  depend on the Lorentz indices and quantum numbers of $J(x)$ and  of the local gauge invariant operators $\hat{O}_N$ built from the quark and gluon fields. These operators are ordered by increasing dimensionality and the Wilson coefficients, calculable in PQCD, fall off by corresponding powers of $-q^2$. In other words, this OPE achieves a factorization of short distance effects encapsulated in the Wilson coefficients, and long distance dynamics present in the vacuum condensates.
Since there are no gauge invariant operators of dimension $d=2$ involving the quark and gluon fields in QCD, it is normally assumed that the OPE starts at dimension $d=4$. This is supported by results from QCD sum rule analyses of $\tau$-lepton decay data, which show no evidence of $d=2$ operators \cite{C2a}-\cite{C2b}.
\begin{figure}[ht]
\begin{center}
\includegraphics[height=.06\textheight]{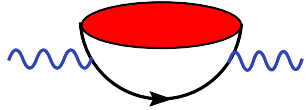}
\caption{\footnotesize{Quark propagator modification due to (infrared) quarks interacting with the physical QCD vacuum, and involving the quark condensate. Large momentum flows through the bottom propagator.}}
\label{fig:figure2}
\end{center}
\end{figure}
The unit operator $\hat{I}$ in Eq.(12) has dimension $d=0$ and $C_0 \hat{I}$ stands for the purely perturbative contribution. The Wilson coefficients as well as the vacuum condensates depend on the renormalization scale. For light quarks, and for the leading $d=4$ terms in Eq.(12), the $\mu^2$ dependence of the quark mass cancels the corresponding dependence of the quark condensate, so that this contribution is a renormalization group (RG) invariant. Similarly, the gluon condensate is also a RG invariant, hence once determined in some channel these condensates can be used throughout.
The numerical values of the vacuum condensates cannot be calculated analytically from first principles as this would be tantamount to solving QCD exactly.
One exception is that of the quark condensate which enters in the Gell-Mann-Oakes-Renner relation \cite{CA2}, a QCD low energy theorem following from the global chiral symmetry of the QCD Lagrangian \cite{GMOR}. Otherwise, it is possible to extract values for the leading vacuum condensates using QCDSR together with experimental data, e.g. $e^+ e^-$ annihilation into hadrons, and hadronic decays of the $\tau$-lepton. Alternatively, as  LQCD  improves in accuracy it should become a valuable source of information on these condensates.\\
\begin{figure}[ht]
\begin{center}
\includegraphics[height=.12\textheight]{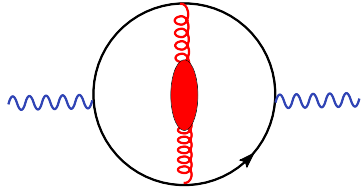}
\caption{\footnotesize{Gluon propagator modification due to (infrared) gluons interacting with the physical QCD vacuum, and involving the gluon condensate. Large momentum flows through the quark propagators.}}
\label{fig:figure3}
\end{center}
\end{figure}

\section{Quark-hadron duality and finite energy QCD sum rules}
Turning to the hadronic sector, bound states and resonances appear in the complex energy (squared) plane (s-plane) as poles on the real axis, and singularities in the second Riemann sheet, respectively. All these singularities lead to a discontinuity across the positive real  axis. Choosing an integration contour as shown in Fig. 4, and given that there are no other singularities in the complex s-plane, Cauchy's theorem leads to the FESR
\begin{figure}[ht]
\begin{center}
  \includegraphics[height=.25\textheight]{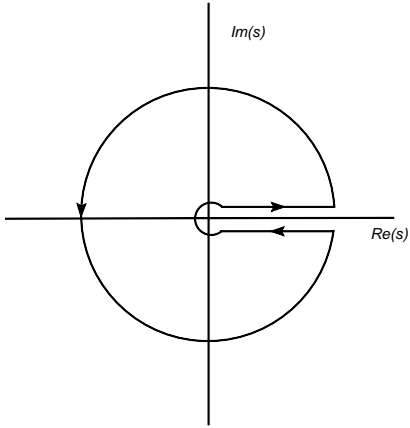}
  \caption{\footnotesize{Integration contour in the complex s-plane. The discontinuity across the real axis brings in the hadronic spectral function, while integration around the circle involves the QCD correlator. The radius of the circle is $s_0$, the onset of QCD.}}
\label{fig:figure4}
\end{center}
\end{figure}
\begin{equation}
\int_{\mathrm{sth}}^{s_0} ds\; \frac{1}{\pi}\; f(s) \;Im \,\Pi(s)|_{HAD} \; = \; -\, \frac{1}{2 \pi i} \; \oint_{C(|s_0|) }\, ds \;f(s) \;\Pi(s)|_{QCD} \; + {\mbox{Res}} [ \Pi(s) f(s), s=0]\;.
\end{equation}
where $f(s)$ is an arbitrary function, $s_{th}$ is the hadronic threshold, and the finite radius of the circle, $s_0$, is large enough for QCD and the OPE to be used on the circle. Depending on the particular form of the integration kernel, $f(s)$, the last term above may or may not be present.
Physical observables determined from FESR should be independent of $s_0$. In practice, though, 
this  is not exact, and there is usually a region of stability where
observables are fairly independent of $s_0$, typically somewhere inside the range $s_0 \simeq 1 - 4 \; \mbox{GeV}^2$. Since $f(s)$ is often a polynomial, the existence of a wide stability region is a highly non-trivial feature.
Equation (13) is the mathematical statement of what is usually referred to as quark-hadron duality. Since QCD is not valid in the time-like region ($s \geq 0$), in principle there is a possibility of problems on the circle near the real axis (duality violations). This issue was identified very early in \cite{Shankar} long before the present formulation of QCDSR, and is currently referred to as (quark-hadron) duality violations. First attempts at identifying and quantifying this problem were made in \cite{PINCH1} using data on hadronic decays of the tau-lepton, the pseudoscalar (pionic) channel and the strangeness changing scalar channel, and in \cite{PINCH2} by considering chiral sum rules. It appears that the size of these duality violations might be channel dependent, as an analysis of tau-decay data extended beyond the kinematical end point finds no effect \cite{FPI}. For recent work on this problem see e.g. \cite{PINCH3}-\cite{PINCH4}, and references therein.\\ 

The right hand side  of this FESR involves the QCD correlator which is expressed in terms of the OPE as in Eq.(12). The left hand side involves the hadronic spectral function which, in principle, is written as
\begin{equation} 
Im \,\Pi(s)|_{HAD}\,=\, Im \,\Pi(s)|_{POLE}\,+\, Im \,\Pi(s)|_{RES} \,\theta(s_0-s)\,+\, Im\, \Pi(s)|_{PQCD}\,\theta(s-s_0) \;,
\end{equation}
where the ground state pole (if present) is followed by the resonances which merge smoothly into the hadronic continuum above some threshold $s_0$. This continuum is expected to be well represented by PQCD if $s_0$ is large enough. Hence, if one were to consider an integration contour in Eq.(13) extending to infinity, the cancellation between the hadronic continuum on the left hand side and the PQCD contribution on the right hand side, would render the sum rule a FESR. In practice, though, there is a finite value $s_0$ beyond which this cancellation takes place, and $s_0$ is identified with the onset of PQCD. In this case the last term in Eq.(14) is obviously redundant.
The integration in the complex s-plane of the QCD correlator is usually carried out in two different ways, Fixed Order Perturbation Theory (FOPT), and Contour Improved Perturbation Theory (CIPT). The first method treats running quark masses and the strong coupling as fixed at a given value of $s_0$. After integrating all logarithmic terms ($\ln(-s/\mu^2)$) the RG improvement is achieved by setting the renormalization scale to $\mu^2 = - s_0$. In CIPT the RG improvement is performed before integration, thus eliminating logarithmic terms, and the running quark masses and strong coupling are integrated  around the circle. This requires solving numerically the RGE for the quark masses and the coupling at each point on the circle.
The FESR Eq.(13) with $f(s)=1$ and  in FOPT can be written as
\begin{equation}
(-)^N \, C_{2N+2} \, \langle0| \hat{O}_{2N+2}|0 \rangle =  \int_0^{s_0} \,ds\, s^N \, \frac{1}{\pi}\, Im \,\Pi(s)|_{HAD} \,-\, s_0^{N+1} \; M_{2N+2}(s_0) \, ,
\end{equation}
where the dimensionless PQCD moments $M_{2N+2}(s_0)$ are given by
\begin{equation}
M_{2N+2}(s_0) = \frac{1}{s_0^{(N+1)}} \, \int_0^{s_0}\, ds\,s^N \, \frac{1}{\pi} \, Im \, \Pi(s)|_{PQCD}\;.
\end{equation}
If the hadronic spectral function is known in some channel from experiment, e.g. from $\tau$-decay into hadrons, then $Im \,\Pi(s)|_{HAD} \equiv Im \,\Pi(s)|_{DATA}$, and Eq.(15) can be used to determine the values of the vacuum condensates. Subsequently, Eq.(15) can be used in a different channel for a different application. It is important to mention that the correlator $\Pi(q^2)$ is generally not a physical observable. However, this has no effect in FOPT as the unphysical quantities (polynomials) in the correlator do not contribute to the integrals. In the case of CIPT, though, this requires modified sum rules involving as many derivatives of the correlator as necessary to render it physical.

\section {Laplace transform QCD sum rules}
The original QCD sum rule method proposed in \cite{SVZ} had as a starting point the well known dispersion relation, or Hilbert transform, which follows from Cauchy's theorem in the complex squared energy $s$-plane
\begin{equation}
\frac{1}{N !} \, \left(- \frac{d}{d Q^2}\right)^N \; \Pi(Q^2)|_{Q^2=Q_0^2} \,=\, \frac{1}{\pi} \, \int_0^\infty \frac{Im\, \Pi(s)}{(s+Q_0^2)^{N+1}} \, ds \;,
\end{equation}
where $N$ is the number of derivatives required for the integral to converge asymptotically, and $Q^2 \equiv - q^2 > 0$. As it stands, the dispersion relation, Eq.(17), is a tautology. Next, a specific asymptotic limit process in the parameters $N$ and $Q^2$ was performed, i.e. $\lim Q^2 \rightarrow \infty$ and $\lim N \rightarrow \infty$, with $Q^2/N \equiv M^2$ fixed, leading to Laplace transform QCDSR  
\begin{eqnarray}
 \hat{L}_M [\Pi(Q^2)] &\equiv& \lim_{\stackrel{Q^2,N \rightarrow \infty}{Q^2/N\equiv M^2}} \frac{(-)^N}{(N-1)!}\, (Q^2)^N \left(\frac{d}{d Q^2} \right)^N\; \Pi(Q^2) \equiv \Pi(M^2) \nonumber \\ [.3cm]
 &=& \frac{1}{M^2} \int_0^{\infty} \frac{1}{\pi} \, Im \Pi(s) \; e^{-s/M^2} \; ds \;.
\end{eqnarray}
This equation is still a tautology. In order to turn it into something with useful content one needs to invoke Eq.(14). This procedure makes no explicit use of the concept of quark-hadron duality, thus not relying on Cauchy's theorem in the complex $s$-plane, other than initially at the level of Eq.(17).
In applications of these sum rules $\Pi(M^2)$ was computed in QCD by applying the Laplace operator $\hat{L}_M$ to the OPE expression of $\Pi(Q^2)$,  and the spectral function on the right hand side was parametrized as in Eq.(14). The function $\Pi(M^2)$ in PQCD involves the transcendental function  $\mu(t,\beta,\alpha)$ \cite{Erdelyi}, as first discussed in \cite{EdeR}.
This novel method had an enormous impact, as witnessed by the several thousand publications to date on analytic solutions to QCD in the non-perturbative domain \cite{QCDSR}. However, in the past decade, and as the subject moved towards high precision determinations to compete with  LQCD, this particular sum rule has fallen out of favour for a variety of reasons as detailed next.\\

The first thing to notice in Eq.(18) is the introduction of an ad-hoc new parameter, $M^2$, the Laplace variable, which determines the squared energy regions where the exponential kernel would have a minor/major impact. It has been regularly advertised in the literature that a judicious choice of $M^2$ would lead to an exponential suppression of the often experimentally unknown resonance region beyond the ground state, as well as to a factorial suppression of higher order condensates in the OPE. In practice, though, this was hardly factually achieved, thus becoming an oracular statement. Indeed, since the parameter $M^2$ has no physical significance, other than being a mathematical artifact, results from these QCDSR would have to be independent of $M^2$ in a hopefully broad region. This so called stability window is often unacceptably narrow, and the expected exponential suppression of the unknown resonance region does not materialize. Furthermore, the factorial suppression of higher order condensates only starts at dimension $d=6$ with a mild suppression by a factor $1/\Gamma(3) = 1/2$. But beyond $d=6$ little, if anything, is numerically known about the vacuum condensates to profit from this feature. Another serious shortcoming of these QCDSR is that the role of the threshold for PQCD in the complex $s$-plane,  $s_0$, i.e. the radius of the circular contour in Fig. 4, is often exponentially suppressed, or at best reduced in importance compared with its role in FESR. This is unfortunate, as $s_0$ is a parameter which, unlike $M^2$, has a clear physical interpretation, and which can be easily determined from data in some instances. On a separate issue,  Laplace sum rules, unlike  FESR, do not facilitate the insertion of non-trivial integration kernels. These kernels have been proved essential in the :\\
(a) In modern determinations of the light-quark masses, to quench significantly the experimentally unknown resonance region beyond the pion and kaon poles, and thus reducing considerably systematic uncertainties from this sector (see Section 6). The Laplace exponential kernel is unsuited for this purpose, thus making it close to impossible to eliminate this systematic uncertainty. Hence, quark-mass determinations in this framework are all affected by some 20-30 \% error, a fact hardly acknowledged in the literature.\\
(b) In tuning the contribution of data in the charm- and bottom-quark regions, thus allowing for very high precision determinations of these quark masses (see Section 7).\\
(c) In allowing for a purely theoretical determination of the charm- and bottom-quark region contributions to the hadronic part of the $g-2$ of the muon \cite{g-2}, in excellent agreement with later LQCD results \cite{g-2LQCD}.\\
(d) Similar to (c) but in relation to the hadronic contribution to the QED running coupling at the scale of the Z-boson \cite{alphaem}.\\
Last, but not least, Laplace sum rule results are often too dependent on the renormalization scale $\mu^2$. In fact, in some applications results are linearly dependent on $\mu^2$, with no plateau in sight, i.e. straight lines with large slopes.\\
Many of these shortcomings of the Laplace QCDSR can be traced back to the way  Cauchy's theorem in the complex $s$-plane is being invoked. This is done trivially, and only initially at the level of the dispersion relation, Eq.(17). In contrast, FESR  are derived directly from Cauchy's theorem, with the upper limit of the integration range, i.e. the radius of the contour $s_0$, being finite on account of the quark-hadron duality assumption. If one were to invoke this assumption in the Laplace sum rule, Eq.(18), it would lead to a serious mismatch between the PQCD contribution to right hand side, and its contribution to the left hand side. In fact, in the integration range $s_0 - \infty$ the integral of the PQCD imaginary part has no counterpart in $\Pi(M^2)$ entering the left hand side. The latter involves the transcendental functions $\mu(t,\beta,\alpha)$, while the former does not. For this reason a power series expansion of the exponential in the Laplace sum rules cannot strictly lead to FESR. Clearly, it is still possible to choose the integration kernel $f(s)$ in Eq.(13) to be a negative exponential, thus leading to a different version of Laplace sum rules. However, this would  be very  different from Eq.(18), plus it would still lead to the rest of the shortcomings mentioned above.

\section{Light quark masses}
Traditionally, the light quark masses have been determined using the correlator, Eq.(9), involving  the pseudoscalar currents $J(x) \equiv \partial_\mu A^\mu(x)|^i_j = [\overline{m}_i(\mu) + \overline{m}_j(\mu)]:\overline{q}_j(x) i \gamma_5 q_i(x):$, where $A_\mu(x)$ is the axial vector current of flavours $i$ and $j$, $\overline{m}_i(\mu)$ the quark mass  in the $\overline{MS}$ scheme, $\mu$ the renormalization scale and $q_i(x)$ are the quark fields. An issue of major concern in the past was the presence of logarithmic quark-mass singularities in these correlators. This problem  has been satisfactorily resolved some time ago in \cite{LMS1}-\cite{LMS2}.
These correlators are now known to five-loop order in PQCD \cite{CHET5}, and free of logarithmic quark mass singularities. The Wilson coefficients of  the leading power corrections, i.e. the gluon and the quark condensates, are also known up to two-loop level \cite{G2q2}. Higher dimensional condensates, as well as quark mass corrections of order ${\cal{O}}$$(m_i^4)$ (with respect to the one-loop term) and higher turn out to be negligible. From Cauchy's theorem, Eq.(13), the FESR  to determine the quark masses can be written as
\begin{eqnarray}
\delta_5^{QCD}(s_0) \equiv - \frac{1}{2\pi i}
\oint_{C(|s_0|)}
ds \;\psi_{5}^{QCD}(s)\; \Delta_5(s) &=& \delta_5^{HAD} \equiv 2\; f_P^2 \; M_P^4\; \Delta_5(M_P^2)  \nonumber \\ [.3cm]
&+&
\int_{s_{th}}^{s_0}
ds \;\frac{1}{\pi} \;Im \;\psi_{5}(s)|_{RES}\;\Delta_5(s) \, , 
\end{eqnarray}
where $\Delta_5(s)$ is an (analytic) integration kernel to be introduced shortly,  the first term on the right hand side  is the pseudoscalar meson pole contribution ($P = \pi, K$), $s_{th}$ is the hadronic threshold, and $Im\, \psi_5(s)|_{RES}$ is the hadronic resonance spectral function. The radius of integration $s_0$ is assumed to be large enough for QCD to be valid on the circle. For later convenience this FESR can be rewritten as
\begin{equation}
\delta_5(s_0)|_{QCD} \,=\, \delta_5|_{POLE}\, + \, \delta_5(s_0)|_{RES}\;, 
\end{equation}
where the meaning of each term is self evident. Historically, the problem with the pseudoscalar correlator has been the lack of direct experimental information on the hadronic resonance spectral functions. Two radial excitations of the pion and of the kaon, with known masses and widths, have been observed in hadronic interactions \cite{PDG}. However, this information is hardly enough to reconstruct the full spectral function. In fact, inelasticity, non-resonant background and resonance interference are impossible to guess, leaving no choice but to model these functions. This introduces an unknown systematic uncertainty which has been present in all past QCD sum rule determinations of the light quark masses. Since the FESR Eq.(19) is valid for any analytic  $\Delta_5(s)$ one can choose this kernel in such a way as to suppress $\delta_5(s_0)|_{RES}$ as much as possible. An example of such a function is the second degree polynomial \cite{ss}-\cite{IJMPA}
\begin{equation}
\Delta_5(s)|_{RES} \,=\, 1 \, - a_0 \,s - a_1\, s^2\;, 
\end{equation}
where $a_0$ and $a_1$ are constants fixed by the requirement $\Delta_5(M_1^2) = \Delta_5(M_2^2) =0$, where $M_{1,2}$ are the masses of the first two radial excitations of the pion or kaon. This simple kernel suppresses enormously the resonance contribution, which becomes only a couple of a percent of the pole contribution, and well below the current uncertainty due to the strong coupling. This welcome feature is essentially independent of the model chosen to parametrize the resonances. A practical parametrization consists of two Breit-Wigner forms normalized at threshold according to chiral perturbation theory, as first proposed in \cite{CADCHPT1} for the pionic channel, and in \cite{CADCHPT2} for the kaonic channel (for an alternative parametrization see \cite{KM} and references therein).
Detailed results for $\delta_5(s_0)|_{QCD}$, to five-loop order in PQCD and up to dimension $d=4$ in the OPE, after integrating in FOPT may be found in \cite{ms}.  
In the case of CIPT the FESR must be written in terms of the second derivative of the current correlator. This is in order to eliminate the unphysical first degree polynomial present in $\psi_5(s)$, which unlike the case of FOPT would otherwise contribute to the FESR which then becomes
\begin{eqnarray}
- \frac{1}{2\pi i}
\oint_{C(|s_0|)}
&ds& \psi_{5}^{'' QCD}(s)\,[F(s) - F(s_0)] = 2\; f_P^2 \; M_P^4\; \Delta_5(M_P^2)  \nonumber \\ [.3cm]
&+&
\frac{1}{\pi} \; \int_{s_{th}}^{s_0}
ds \; Im \;\psi_{5}(s)|_{RES}\;\Delta_5(s) \, , 
\end{eqnarray}
where
\begin{equation}
F(s) = - s \left(s_0 - a_0\,\frac{s_0^2}{2} - a_1\, \frac{s_0^3}{3} \right) + \frac{s^2}{2} - a_0\, \frac{s^3}{6} - a_1\, \frac{s^4}{12} \;,
\end{equation}
and
\begin{equation}
F(s_0) = - \frac{s_0^2}{2} +  a_0\, \frac{s_0^3}{3} + a_1\, \frac{s_0^4}{4} \;.
\end{equation}
\begin{figure}[ht]
\begin{center}
  \includegraphics[height=.35\textheight]{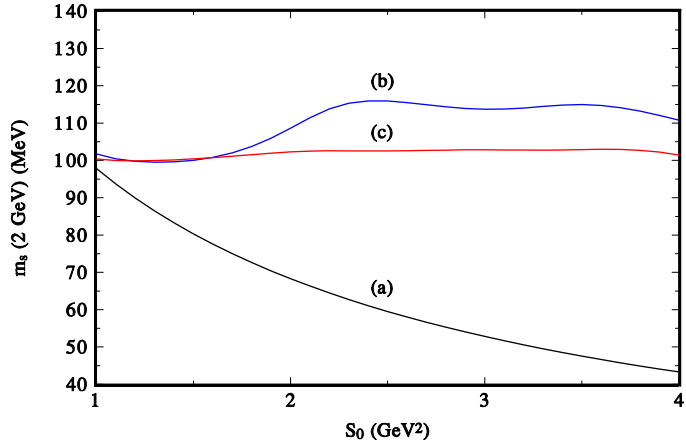}
  \caption{\footnotesize{The strange quark mass $\overline{m}_s (2\; \mbox{GeV})$ in the $\overline{MS}$ scheme taking into account only the kaon pole with $\Delta_5(s) = 1$ (curve(a)), and the two Breit-Wigner resonance spectral function with a threshold constraint from CHPT \cite{CADCHPT2}, with $\Delta_5(s) = 1$ (curve(b)), and $\Delta_5(s)$ as in Eq.(21) (curve(c)). A systematic uncertainty of some $20 \,\%$  due to the resonance sector is dramatically unveiled.}}
  \label{fig:figure7}
  \end{center}
\end{figure} 
The RG improvement is used before integration, so that all logarithmic terms vanish. The running coupling as well as the running quark masses are no longer frozen as in FOPT, but must be integrated. This can be done by solving numerically the respective RG equations at each point on the integration circle in the complex s-plane. Detailed expressions are given in \cite{ms}-\cite{mq}.\\
The parameters of the integration kernel, Eq.(21), are $a_0 = 0.897 \;\mbox{GeV}^{-2}$, and $a_1 = -\, 0.1806 \;\mbox{GeV}^{-4}$ for the pionic channel, and $a_0 = 0.768 \;\mbox{GeV}^{-2}$, and $a_1 = -\, 0.140 \;\mbox{GeV}^{-4}$ for the kaonic channel. These values correspond to the radial excitations $\pi(1300)$, $\pi(1800)$, $K(1460)$ and $K(1830)$. The pion and kaon decay constants are \cite{PDG} $f_\pi = 92.21 \,\pm 0.14 \;\mbox{MeV}$, and $f_K = (1.22 \pm 0.01) f_\pi$. In the QCD sector it is best to use the value of the strong coupling determined at the scale of the $\tau$-mass, as this is close to the scale in current use for the light quark masses, i.e. $\mu= 2 \;\mbox{GeV}$. The extraction of $\alpha_s(M_\tau)$ from the $R_\tau$ ratio  involves an integral with a natural kinematical integration kernel that eliminates the contribution of the $d=4$ term in the OPE. This welcome feature improves the accuracy of the determination, and it makes little sense to introduce additional spurious integration kernels which would artificially recover this $d=4$ contribution. The different values obtained from $\tau$ decay using CIPT are all in agreement with each other, i.e. $\alpha_s(M_\tau) = 0.338 \pm 0.012$ \cite{PICH2}, $\alpha_s(M_\tau) = 0.341 \pm 0.008$ \cite{CVETIC}, $\alpha_s(M_\tau) = 0.344 \pm 0.009$ \cite{DAVIER}, $\alpha_s(M_\tau) = 0.332 \pm 0.016$ \cite{BAIKOV}, and the most recent result $\alpha_s(M_\tau) = 0.331 \pm 0.013$ \cite{Pich14}. These determinations are model independent and
extremely transparent, with $\alpha_s$ obtained essentially by confronting PQCD with the single experimental number $R_\tau$. The $d=4$ gluon condensate has been extracted from $\tau$ decays \cite{C2b}, but one can conservatively consider the wide range$<\alpha_s G^2> = 0.01 - 0.12 \;\mbox{GeV}^4$. The impact of the light quark condensate is at the level of $1 \%$ in the quark masses. A $\pm \;30 \%$ uncertainty in the resonance contribution $\delta_5(s_0)|_{RES}$ in Eq.(20) translates into a safe $1 \%$ change in the quark masses. Finally, it has been assumed that the unknown six-loop PQCD contribution is equal to the five-loop result, a supposedly conservative estimate of higher orders in PQCD.  Nevertheless, there is a convergence problem with the PQCD expansion to be discussed later.\\ 
\begin{table}
\small
\begin{tabular}{ccccccccc}
\hline \\
\noalign{\smallskip}
Source & $\overline{m}_u$ & $\overline{m}_d$& $\overline{m}_s$& $\overline{m}_{ud}$ & $\overline{m}_u/\overline{m}_d$ & $\overline{m}_s/\overline{m}_{ud}$ & $R$ & $Q$ \\
\hline \\
\noalign{\smallskip}
QCDSR \cite{mq}  &  $2.9 \pm 0.2$ & $5.3 \pm 0.4$ &  - & $4.1 \pm 0.2$   &  $0.553$ & -  &- &  -\\
&&&&& (input) &&&\\
\hline \\
\noalign{\smallskip}
QCDSR \cite{ms}  &  - & - & $ 102 \pm 8$  & -  & - & $24.9 \pm 2.7$  &$33 \pm 6$ & $21 \pm 3$ \\
\hline \\
QCDSR \cite{ms13}  &  - & - & $94 \pm 9$  & -  & - & $27 \pm 1$  &$30 \pm 6$ & $19 \pm 3$ \\
& & & &&& (input)&&\\
\hline \\
\noalign{\smallskip}
FLAG \cite{FLAGR}  &  $2.40 \pm 0.23$ &  $4.80 \pm 0.23$  & $ 101 \pm 3$ &  $3.6 \pm 0.2$   & $0.50 \pm 0.04$  & $28.1 \pm 1.2$  & $40.7 \pm 4.3$ & $24.3 \pm 1.5$ \\
\noalign{\smallskip}
\hline
\end{tabular}
\caption{\footnotesize{The running quark masses in the $\overline{MS}$ scheme at a scale $\mu = 2 \; \mbox{GeV}$ in units of MeV from QCDSR (first three rows), and from the FLAG lattice QCD analysis \cite{FLAGR}. The ratios $\overline{m}_u/\overline{m}_d$ and  $\overline{m}_s/\overline{m}_{ud}$ are an input in the QCDSR (see text). The ratios $R$ and $Q$ are defined on the l.h.s. of  Eqs.(2) and (3).}}
\end{table}
\normalsize
Beginning with the strange quark mass,  Fig. 1 shows the results for $\overline{m}_s (2\; \mbox{GeV})|_{\overline{MS}}$ with no integration kernel, $\Delta_5(s) = 1$, and taking into account only the kaon pole, curve (a), and the kaon pole plus a two Breit-Wigner resonance model with a threshold constraint from CHPT \cite{CADCHPT2}, curve (b) (a misprint in the formula for the spectral function in \cite{CADCHPT2} has been corrected in  \cite{CADCHPT3}). These curves are for the central value of $\alpha_s(M_\tau)$ whose uncertainties will be considered afterwards.
The latter result is reasonably stable in the wide region $s_0 = 2 - 4 \; \mbox{GeV}^2$, so that it could lead us to conclude that $\overline{m}_s (2\; \mbox{GeV})|_{\overline{MS}}  \simeq 100 - 120 \; \mbox{MeV}$, albeit with a yet unknown systematic uncertainty arising from the resonance sector.
Introducing the kernel, Eq.(21), leads to curve (c) and to a dramatic unveiling of this systematic uncertainty. In fact, the {\it real} value of the quark mass is $\overline{m}_s (2\; \mbox{GeV})|_{\overline{MS}} = 102 \pm 8 \; \mbox{MeV}$, or some $20 \%$ below the former result (this error now includes the uncertainty in $\alpha_s$). In addition, and as a bonus, the systematic uncertainty-free result is remarkably stable in the  unusually wide region $s_0  \simeq 1 - 4 \; \mbox{GeV}^2$ (typical stability regions are only half as wide).\\ It must be recalled that the pseudoscalar correlator involves the overal factor $(m_s + m_{ud})^2$. Hence, in order to determine $m_s$ an input value for the ratio $m_s/m_{ud}$ is needed in the result from the sum rule, which is
\begin{equation}
\overline{m}_s (2\; \mbox{GeV})|_{\overline{MS}} = \frac{105.5 \pm 8.2 \; \mbox{MeV}}{1 + m_{ud}/m_s} \;.
\end{equation}
Using the wide range $m_s/m_{ud} = 24 - 29$ leads to $\overline{m}_s (2\; \mbox{GeV})|_{\overline{MS}} = 102 \pm 8 \; \mbox{MeV}$. In this case the impact of the uncertainty in the quark mass ratio is  small.
 However, in the case of the up- and down-quark masses  the corresponding ratio $m_u/m_d$ plays a more important role in the result from the sum rule, which is
\begin{equation}
\overline{m}_d (2\; \mbox{GeV})|_{\overline{MS}} = \frac{8.2 \pm 0.6 \; \mbox{MeV}}{1 + m_u/m_d} \;.
\end{equation}
The input used in \cite{mq}  for the mass ratio in Eq.(26) is $m_u/m_d = 0.553$   from CHPT \cite{LEUT}. Once $m_d$ is determined from Eq.(26),  $m_u$ follows. Using these results for the individual masses one obtains the ratios $m_u/m_d$ and $m_s/m_{ud}$ shown in Table 1. Using instead the ratio $m_u/m_d = 0.50 \pm 0.04$ from \cite{FLAGR} gives $m_u = 2.7 \pm 0.3 \; {\mbox{MeV}}$, and $m_d = 5.5 \pm 0.4 \; {\mbox{MeV}}$. These quark masses $\overline{m}_u (2\; \mbox{GeV})|_{\overline{MS}}$ and $\overline{m}_d (2\; \mbox{GeV})|_{\overline{MS}}$
also exhibit a remarkably wide stability region $s_0 \simeq 1 - 4 \; \mbox{GeV}^2$ \cite{mq},\cite{IJMPA}.\\

I now return to the problem with the PQCD convergence of the pseudoscalar correlator, and its impact on the strange-quark mass \cite{ms13} (the impact on the up- and down-quark masses, leading to improved results, is currently under investigation \cite{mud}). For the integration kernel, Eq.(21), and $s_0= 4.2 \; {\mbox{GeV}}^2$, the FOPT result for $\delta_5^{QCD}(s_0)$ in PQCD is given by
\begin{equation}
\delta_5^{PQCD}= 0.23 \, {\mbox{GeV}}^{8} \left[1 + 2.2 \,\alpha_{s} + 6.7\,\alpha_{s}^2 + 19.5 \,\alpha_{s}^3 + 56.5\, \alpha_{s}^4\right]\;,
\end{equation} 
which after replacing a typical value of $\alpha_s$ leads to all terms beyond the leading order to be roughly the same, e.g. for $\alpha_s = 0.3$ the result is
\begin{equation}
\delta_5^{PQCD}= 0.23 \, {\mbox{GeV}}^{8} \left[1 + 0.65 + 0.60 + 0.53 + 0.46\right]\;,
\end{equation}
which is hardly (if at all) convergent. In fact, judging from the first five terms, this expansion
is worse behaved than the non-convergent harmonic series. An integration kernel \cite{ms13} shown to be better suited than Eq.(21) is  
\begin{equation}
 \Delta_5(s) = (s-a)(s-s_0) \;,
\end{equation}
with $a=2.8\; {\mbox{GeV}^2}$. In this case the perturbative expansion for the strange-quark mass, with $m_s/m_{ud} = 27 \pm 1$, becomes
\begin{equation}
\overline{m}_s(2 \,{\mbox{GeV}})= 248.3\; {\mbox{MeV}} (1 + 2.59\, \alpha_s + 8.60\, \alpha_s^2 + 26.50\, \alpha_s^3 + 75.47\, \alpha_s^4)^{-1/2}\;, 
\end{equation}
with all terms being roughly of the same size for $\alpha_s(2\,{\mbox{GeV}}) \simeq 0.3$. This result implies an obvious systematic uncertainty in the QCD sector, which was not exposed and dealt with in \cite{ms}. The ideal tool to deal with this problem is that of Pad\`{e}  approximants \cite{Pade}
\begin{equation}
f(z) \approx  [m/n] \equiv \frac{a_{0} + a_{1} z + ... + a_{m}  z^m}{1 + b_1 z + ... + b_n z^n} \,, \;\;\; m+n = k \;,
\end{equation}
with $[m/0]$ being the standard Taylor series expansion of $f(z)$. This particularly simple Pad\`{e} approximant already accelerates the PQCD convergence 
\begin{equation}
\overline{m}_s(2 \,{\mbox{GeV}})= 248.3\; {\mbox{MeV}} (1 - 1.30\, \alpha_s + 1.80\, \alpha_s^2 - 1.95\, \alpha_s^3 - 0.34\, \alpha_s^4)^{-1/2}\;, 
\end{equation}
leading to
\begin{equation}
\overline{m}_s (2\; \mbox{GeV})|_{\overline{MS}} = \frac{97.48 \pm 10.6 \; \mbox{MeV}}{1 + m_{ud}/m_s} = 94 \, \pm 9 \,{\mbox{MeV}}  \;.
\end{equation}
In contrast, the original expression, Eq.(30) would give $\bar{m}_s( 2\;{\mbox{GeV}}) \simeq 125 \,{\mbox{MeV}}$. A systematic uncertainty in the QCD sector of roughly 30\% has thus been  exposed and eliminated\!. For a more detailed discussion of this procedure see \cite{SBPhD}.\\

In Table 1  one  finds a summary of the results for the light quark masses, and the ratios $R$ and $Q$ defined on the left hand side of Eqs. (2) and (5), together with the results of the Flag group \cite{FLAGR}. The values of the up- and down-quark masses and their ratios in Table 1 are slightly different from those in \cite{IJMPA} due to the input value for the ratio $m_u/m_d$. Using the FLAG ratio \cite{FLAGR} instead, gives similar values within errors as mentioned earlier after Eq.(26). In either case there seems to be some tension between these results and those from \cite{FLAGR}. Perhaps once the QCD systematic uncertainty is dealt with in this channel the tension might be resolved. \\
The various sources of errors in the quark masses discussed earlier combine into the final values given in Table 1. Having all but eliminated the systematic uncertainty from the hadronic resonance sector,  the main source of error is now due to the strong coupling, and the PQCD sector for the up- and down-quark masses, i.e. the poor convergence of the perturbative series. Improved accuracy in the determination of $\alpha_s$ would then allow for a reduction of the uncertainties in the light quark mass sector.
\section{Heavy quark masses}
Determinations of the charm- and bottom-quark masses are not affected by a lack of data, as there is plenty of experimental information from $e^+ e^-$ annihilation into hadrons at high energies \cite{PDG}, except for a gap in the region
$25 \;\mbox{GeV}^2 \lesssim s \lesssim 50\; \mbox{GeV}^2$ . On the theoretical side there has been very good progress on PQCD  up to four-loop level \cite{QCD1}-\cite{QCD14}. The leading power correction in the OPE is due to the gluon condensate with its Wilson coefficient known at the two-loop level \cite{BROAD}.
\begin{table}
\begin{center}
\begin{tabular}{ccccc}
\hline \\
\multicolumn{4}{r}{$\overline{m}_c(3\; \mbox{GeV})$(in MeV)} \\
\cline{2-5}
\noalign{\smallskip}
 Kernel  & $\bar{m}_{c}^{(0)}$ &$\bar{m}_{c}^{(1)}$  & $\bar{m}_{c}^{(2)}$ &$\bar{m}_{c}^{(3)}$   \\
\hline
\noalign{\smallskip}
$s^{-2}$ & 1129 & 1021 & 998 & 995  \\
$1-(s_0/s)^2$ & 1146 & 1019 & 991 & 987    \\
\hline\\
\end{tabular}
\caption{\footnotesize{Results for the charm-quark mass at different orders in PQCD, and for two integration kernels from \cite{SB2}. The result for $f(s) = 1/s^2$ is obtained using slightly different values of the QCD parameters, and a different integration procedure as in \cite{{Kuhnmc}}.}}
\end{center}
\end{table}
%
\begin{table}
\begin{center}
\begin{tabular}{ccccccc}
\hline \\
\multicolumn{6}{r}{Uncertainties (in MeV)} \\
\cline{3-7}
\noalign{\smallskip}
 Kernel & $\bar{m}_c(3\,\mbox{GeV})$ & Exp. & $\Delta \alpha_s$  & $\Delta \mu$ & NP &    Total               \\
\hline
\noalign{\smallskip}
$s^{-2}$  &  995 \quad & \quad 9 \quad &\quad  3 \quad &\quad  1 \quad  &\quad  1 \quad &\quad 9.6\\
$1-(s_0/s)^2$  &  987 \quad & \quad 7 \quad &\quad  4 \quad &\quad  1 \quad  &\quad  1 \quad &\quad 8.2\\
\hline\\
\end{tabular}
\caption{\footnotesize{The various uncertainties due to the data (EXP), the value of $\alpha_s$ ($\Delta \alpha_s$), changes of $\pm 35 \%$ in the renormalization scale around $\mu= 3 \; \mbox{GeV}$ ($\Delta \mu$), and the value of the gluon condensate (NP) \cite{SB2}.}}
\end{center}
\end{table}
The correlator, Eq.(9), involves the vector current $J(x)\equiv V_\mu(x) = \bar{Q}(x) \gamma_\mu Q(x)$, where $Q(x)$ is the charm- or bottom-quark field. The experimental data is in the form of the $R_Q$-ratio for charm (bottom) production, which determines the hadronic spectral function. Modern determinations of the heavy-quark masses have been based on inverse moment (Hilbert-type) QCDSR, e.g. Eq.(13) with $f(s) = 1/s^n$, in which case  Eq.(13) requires the additional term on the right hand side, i.e. the residue at the pole: ${\mbox{Res}} [ \Pi(s) f(s), s=0]$. These sum rules require QCD knowledge of the vector correlator in the low energy region, around the open charm (bottom) threshold, as well as in the high energy region. A recent update \cite{KUHN10} of earlier determinations \cite{QCD1}-\cite{QCD3}, \cite{QCD5}-\cite{QCD7} reports a charm-quark mass in the $\overline{MS}$ scheme accurate to $1 \%$, and half this uncertainty for the bottom-quark mass. However, the analysis of \cite{QCD14} claims an error a factor two larger for the charm-quark mass. It appears that the discrepancy arises from the treatment of PQCD. In fact, in \cite{QCD14} two different renormalization scales were used, one for the strong coupling and another one for the quark mass. This unconventional choice results in an artificially  larger error in the charm-quark mass obtained from  inverse (Hilbert) moment QCDSR. It does not affect, though, sum rules involving positive powers of $s$. In any case, the philosophy in current use is to choose the result from the method leading to the smallest uncertainty. \\ 
Beginning with the charm-quark mass, an alternative procedure was proposed some years  ago based  only on the high energy expansion of the heavy-quark vector correlator \cite{KS1}-\cite{KS2}. This method was followed recently \cite{SB1}, but with updated PQCD information and the inclusion of  integration kernels in the FESR, Eq.(13), tuned to enhance/suppress contributions from data in certain regions. One  such kernel is  the so-called {\it pinched} kernel \cite{PINCH1}-\cite{PINCH2}
\begin{equation}
f(s) = 1 - \frac{s}{s_0} \;,
\end{equation}
which is supposed to suppress potential duality violations close to the real s-axis in the complex s-plane. In connection with the charm-quark mass application, this kernel enhances the contribution from the first two narrow resonances, $J/\psi$ and $\psi(2S)$, and reduces the weight of the broad resonance region, particularly near the onset of the continuum. 
The latter feature is better achieved with the alternative kernel \cite{SB2}
\begin{equation}
f(s) = 1 - \left(\frac{s_0}{s}\right)^2 \;,
\end{equation}
which produces an obvious larger enhancement of the narrow resonances, and a higher quenching of the broad resonance region. This kernel has been used together with both the high and the low energy expansion of the vector correlator in \cite{SB2}. 
The results for $\overline{m}_c(3\,\mbox{GeV})$ in the $\overline{MS}$ scheme using two different integration kernels are listed in Table 2, and the related uncertainties are shown in Table 3. The kernel $f(s) = 1/s^2$ is from \cite{Kuhnmc}. The merits of each kernel may be judged by its ability to minimize these uncertainties, in particular those that might be most affected by systematic errors, such as  e.g. the experimental data. The kernel Eq.(35) appears to be optimal as it produces the smallest uncertainty due to the data, and is very stable against changes in $s_0$. Some recent determinations of the charm-quark mass  are based on Hilbert moments with no  $s_0$-dependent kernel, such as that in Table 2. While  there is no explicit $s_0$-dependence in Hilbert moments (the integrals extend to infinity), there is definitely a residual dependence when choosing the threshold for the onset of PQCD. From a FESR perspective, the major drawback of the kernel $1/s^2$ is clearly the poor stability against changes in $s_0$. An important remark is in order concerning the uncertainty due to changes in $s_0$. From current data it is not totally clear where does PQCD actually start. This problem not only affects FESR, with their explicitly obvious $s_0$-dependence, but also Hilbert moments with an implicit $s_0$-dependence, as there is no data all the way up to infinity.
\begin{table}[ht!]
\begin{center}
\small
\begin{tabular}{lccccccccc}
\hline\\
& &  & \multicolumn{4}{c}{Uncertainties (MeV)} &  \multicolumn{3}{c}{\textbf{Options A, B, C} (MeV)}\\
\cline{4-7}
\cline{8-10}
\noalign{\smallskip}
$f(s)$ 	& $\overline{m}_b(10\,\mbox{GeV})$ & $\sqrt{s_0}\,(\text{GeV})$  & $\Delta\text{EXP.}$ & 	$\Delta \alpha_s$  &	 $\Delta \mu$			&	 $\Delta$TOTAL	  &  $\Delta\textbf{A}$&$\Delta\textbf{B}$& $\Delta\textbf{C}$\\
\hline
\noalign{\smallskip}
$s^{-3}$ 			&  	3612  						&  $\infty$		&		9	 		&		4	 	&		1	 &	10		 & 		20	 &	-17 & 16 \\
$s^{-4}$ 			&  	3622  						&  $\infty$		&		7	 		&		5	 	&		10	 &	13		 & 		12	 &	-12 & 8 \\ \\
$\mathcal{P}_{3}^{(-3,-1,0)}(s_0,s)$ 	&  	 3623 		& 16			&   6		 &		6	 	&		2	 &	9		 & 		1	 &	-6 & 0 \\
$\mathcal{P}_{3}^{(-3,-1,1)}(s_0,s)$ 	&  	 3623 		& 16			&   6		 &		6	 	&		2	 &	9		 & 		2	 &	-7 & 0 \\
$\mathcal{P}_{3}^{(-3,0,1)}(s_0,s)$ 	&  	 3624 		& 16			&   7		 &		6	 	&		2	 &	9		 & 		2	 &	-7 & 0 \\
$\mathcal{P}_{3}^{(-1,0,1)}(s_0,s)$ 	&  	 3625 		& 16			&   8		 &		5	 	&		4	 &	10		& 	4	 &	-12 & 0 \\ \\
$\mathcal{P}_{4}^{(-3,-1,0,1)}(s_0,s)$ 	&  	 3623 		& 20			&   6		 &		6	 	&		3	 &	9		& 	0	 &	-4 & 0 \\
\hline\\
\end{tabular}
\caption{\footnotesize{Results from \cite{bquark} for $\overline{m}_b(10\,\mbox{GeV})$ using kernels $f(s)$ selected for producing the lowest uncertainty. Results from the kernels $f(s)=s^{-3}$ and $f(s)=s^{-4}$ used in \cite{KUHN10},\cite{Chetet1} are given here for comparison. The errors are from experiment \cite{BABAR}, ($\Delta\text{EXP.}$), from the strong coupling ($\Delta \alpha_s$) and from variation of the renormalization scale by $\pm \, 5 \, \mbox{GeV}$ around $\mu = 10\, \mbox{GeV}$ ($\Delta \mu$). These sources were added in quadrature to give the total uncertainty ($\Delta$TOTAL). Options {\bf A}, {\bf B}, {\bf C} refer to different ways of treating the data (see \cite{bquark}) . The option uncertainties $\Delta \textbf{A}$, $\Delta \textbf{B}$ and $\Delta \textbf{C}$ are the differences between $\overline{m}_b(10\,\mbox{GeV})$ obtained with and without  \textbf{Option A},  \textbf{B},  or \textbf{C}. As in  \cite{KUHN10},\cite{Chetet1} these are not added to the total uncertainty, and are listed only for comparison purposes.}}\label{Tab:results}
\end{center}
\end{table}

Two of the most recent results for $\overline{m}_c(3\,\mbox{GeV})$ \cite{KUHN10}, \cite{QCD14}, together with the weighted FESR value \cite{SB2}  are 
\begin{eqnarray}
\overline{m}_c(3\,\mbox{GeV}) \;  = \;\Bigg\{ 
\begin{array}{lcl}
986  \; \pm 13 \;\mbox{MeV} \; \cite{KUHN10}\\
998  \; \pm 29 \;\mbox{MeV} \; \cite{QCD14} \\
987  \; \pm \;\; 9 \;\mbox{MeV} \; \cite{SB2}\;, 
\end{array}
\end{eqnarray}
in very good agreement with each other, except for the errors. The small uncertainty from \cite{SB2} is due in part to improved quenching of the data in the broad resonance region, but mostly due to a strong reduction in the sensitivity to $s_0$, i.e.  the onset of PQCD.  For comparison, a recent LQCD determination gives \cite{LATT1}
\begin{equation}
\overline{m}_c(3\,\mbox{GeV}) \;  = \; 986 \; \pm\; 6 \;\mbox{MeV}\;,
\end{equation}
in excellent agreement in magnitude and uncertainty with \cite{SB2}.\\

Turning to the bottom-quark mass, the most recent precision determination   \cite{bquark} is based on $e^+ e^-$ data from the {\it{BABAR}} Collaboration \cite{BABAR}, and  integration kernels related to Legendre-type Laurent polynomials, as used e.g. in the charm-quark case \cite{SB1}, to wit
\begin{equation}\label{intkernel}
 f(s) \equiv \mathcal{P}_{3}^{(i,j,k)}(s,s_0)=A(s^{i}+B s^j+C s^k)\;, 
\end{equation} 
subject to the global constraint
\begin{equation}\label{constraint}
\int_{s^*}^{s_0} \mathcal{P}_{3}^{(i,j,k)}(s,s_0)\; s^{-n}\;ds= 0,
\end{equation} 
where $n\in \{0,1\}$, $i,j,k\in \{-3,-2,-1,0,1\}$, and $i,j,k$ are all different. The above constraint determines the constants $B$ and $C$. The constant $A$ is an arbitrary overall normalization which cancels out in the sum rule Eq.(13). The reason for the presence of the integrand $s^{-n}$ above is that the behaviour of $R_b(s)$ in the region to be quenched resembles a monotonically decreasing logarithmic function. Hence, an inverse power of $s$ optimizes the quenching.
\begin{figure}
\begin{center}
\includegraphics[scale=0.67]{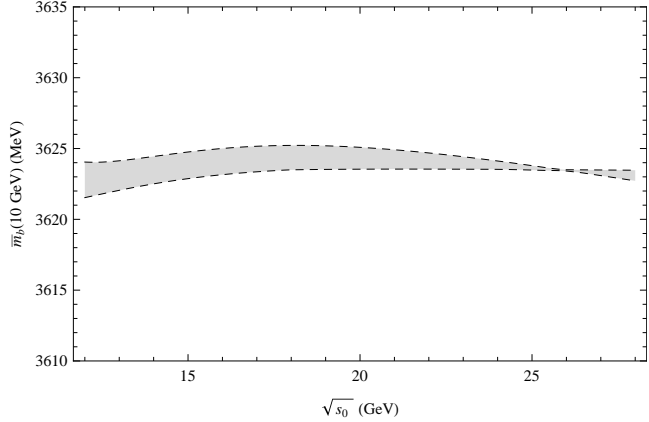}
\caption{\footnotesize{The values of $\overline{m}_b(10\,\mbox{GeV})$, obtained for different values of $s_0$ and using the 10 different kernels in the class $\mathcal{P}_{3}^{(i,j,k)}(s_0,s)$. All results lie within the shaded region.}}\label{Fig:stab1}
\end{center}
\end{figure}
As an example, taking $s_0= (16\,\text{GeV})^2$ (and $A=1$) one finds
\begin{equation}
\mathcal{P}_{3}^{(-3,-1,0)}(s,s_0)= s^{-3} - (1.02\times 10^{-4} \;\mbox{GeV}^{-4}) \;s^{-1} 
+  3.70\times 10^{-7}\; \mbox{GeV}^{-6} \;,
\end{equation}
with $s$ in units of $\mbox{GeV}^2$.
There are  five different kernels $\mathcal{P}_{3}^{(i,j,k)}$, and  the spread of values obtained for $\overline{m}_b$ using this set of different kernels was  used as a consistency check on the method. The fourth-order Laurent polynomial $\mathcal{P}_{4}^{(i,j,k,r)(s,s_0)}$ is also defined by the constraint Eq.(39), but with $n\in {0,1,2}$, and there are also five different kernels of this type.\\
Table 4 summarizes the results obtained in \cite{bquark}. Options {\bf A}, {\bf B}, and {\bf C} refer to different options for considering the data (for details see \cite{bquark}). In \cite{bquark} a total of 15 different kernels, $f(s)$ in Eq.(13), were considered, i.e. 10 from the class $\mathcal{P}_{3}^{(i,j,k)}(s,s_0)$ , and 5 from the class $\mathcal{P}_{4}^{(i,j,k,r)}(s,s_0)$. Figure 6  shows the range of values for $\overline{m}_b(10\,\text{GeV})$ obtained using all of the  10 kernels in the class $\mathcal{P}_{3}^{(i,j,k)}(s,s_0)$, as a function of  $s_0$. Remarkably, between $12\,\text{GeV}<\sqrt{s_0}<28\,\text{GeV}$, all of the masses obtained using all 10 kernels from the class $\mathcal{P}_{3}^{(-3,-1,0)}(s,s_0)$ fall in the range $3621\,\text{MeV}\leq\bar{m}_b(10\,\text{GeV})\leq 3625\,\text{MeV}$. The method gives a consistent result even in the region $\sqrt{s_0}< 4 \overline{m}_b (\mu)\approx 15\,\text{GeV}$ where the high-energy expansion used in the contour integral in Eq.(13) is not guaranteed to converge. Using, rather, the 5 kernels in the class $\mathcal{P}_{4}^{(i,j,k,r)}(s,s_0)$, and varying $s_0$ in the range $18\,\text{GeV}<\sqrt{s_0}<70\,\text{GeV}$, all of the masses thus obtained lie in the interval $3620\,\text{MeV}\leq\bar{m}_b(10\,\text{GeV})\leq 3626\,\text{MeV}$. These results show a great insensitivity of this method on the  parameter $s_0$, and also on which powers of $s$ are used to construct $\mathcal{P}_{3}^{(i,j,k)}(s,s_0)$ and $\mathcal{P}_{4}^{(i,j,k,r)}(s,s_0)$. This in turn demonstrates the consistency between the high and low energy expansions of PQCD.  The final result  chosen in \cite{bquark} from the optimal kernel $\mathcal{P}_{3}^{(-3,-1,0)}(s_0,s)$ is
\begin{equation}
\overline{m}_b(10\,\text{GeV})= 3623(9)\,\text{MeV} \;,  \label{final}
\end{equation}
\begin{equation}
\overline{m}_b(\overline{m}_b)= 4171(9)\,\text{MeV} \;.  \label{final2}
\end{equation}
This value  is fully consistent with the latest lattice determination $\overline{m}_b(10\,\text{GeV})= 3617(25)\,\text{MeV}$ \cite{LATT1}. It is also consistent with a previous QCD sum rule precision determination \cite{Chetet1} giving $\overline{m}_b(10\,\text{GeV})= 3610(16)\,\text{MeV}$.

\section{Conclusions}
After a short review of quark mass ratios the method of QCDSR was discussed,  in connection with determinations of individual values of the quark masses. The historical (unknown) hadronic systematic uncertainty affecting light quark mass determinations for over thirty years was highlighted. Details of the recent breakthrough in strongly reducing this uncertainty were provided. In addition it was explained how the previously unknown systematic QCD uncertainty, due to the poor convergence of the light-quark pseudoscalar correlator, was exposed and essentially eliminated for the strange-quark mass. The up- and down-quark cases are currently under investigation, and should lead to a similar successful result. Future improvement in accuracy is now possible, and depends essentially on more accurate determinations of the strong coupling, the remaining main source of error. In the heavy quark sector recent high precision determinations of the charm- and bottom-quark masses were reported. While these values are all in agreement, there is some disagreement on the size of the errors. The use of suitable multi-purpose integration kernels in FESR allows to tune the weight of the various contributions to the quark masses. This in turn allows to minimize the error due to the data, as well as to the uncertainty in the onset of PQCD. The latter uncertainty impacts FESR as well as Hilbert moment sum rules, as there is no data all the way up to infinity. If no kernel, other than simple (non-pinched) inverse powers of $s$ are used then this uncertainty would be much larger than normally reported, as may be appreciated from Tables 3, and especially 4. However, according to  current philosophy one chooses the determination having the smallest error.
Marginal improvement of the current total error in this framework should be possible with improved accuracy in the data and in the strong coupling.
 
\section{Acknowledgements}
The author wishes to thank his collaborators  in the various projects on quark masses reported here: S. Bodenstein, J. Bordes, N. Nasrallah, J. Pe\~{n}arrocha, R. R\"{o}ntsch and  K. Schilcher. This work was supported in part by NRF (South Africa).
 

\end{document}